\documentclass[11pt]{article}

\usepackage[compress]{cite}

\usepackage{amsfonts,amsthm}
\usepackage{amsmath}  
\usepackage{latexsym}
\usepackage{amssymb}
\usepackage{mathrsfs}
\usepackage{tikz}
\usepackage[all]{xy}

\newtheorem{thm}{Theorem}[section]
\newtheorem{lem}[thm]{Lemma}
\newtheorem{prop}[thm]{Proposition}

\newtheorem{coro}[thm]{Corollary}

\newtheorem{rem}[thm]{Remark}
\newtheorem{exa}[thm]{Example}

\newtheorem*{term*}{Notation/Terminology}

\newtheorem{defi}[thm]{Definition}

\newcommand{\si}{\sigma}

\newcommand{\CC}{\mathbb{C}}

\begin{document}

\title{Baxterisation of the fused Hecke algebra \\ and $R$-matrices with $gl(N)$-symmetry}

\author{N. Cramp\'e\footnote{Institut Denis-Poisson CNRS/UMR 7013 - Universit\'e de Tours - Universit\'e d'Orl\'eans, 
Parc de Grandmont, 37200 Tours, France. crampe1977@gmail.com}$\ $ and 
L. Poulain d'Andecy\footnote{Laboratoire de math\'ematiques de Reims UMR 9008, Universit\'e de Reims Champagne-Ardenne,
Moulin de la Housse BP 1039, 51100 Reims, France.
loic.poulain-dandecy@univ-reims.fr}}

\date{}
\maketitle

\begin{abstract}
We give an explicit Baxterisation formula for the fused Hecke algebra and its classical limit for the algebra of fused permutations. 
These algebras replace the Hecke algebra and the symmetric group in the Schur--Weyl duality theorems for the symmetrised powers of the fundamental 
representation of $gl(N)$ and their quantum version. So the Baxterisation formulas presented in this paper are applicable to the $R$-matrices associated to these representations. 
In particular all ``higher spins'' representations of (classical and quantum) $sl(2)$ are covered.
\end{abstract}

\section{Introduction}

The Yang--Baxter (YB for short) equation with spectral parameters and its solutions, the $R$-matrices, emerged in different contexts of the theoretical physics: 
in quantum mechanics, it appears as a constraint during the resolution by the coordinate Bethe ansatz of the $n$-body problem with $\delta$-function interaction  \cite{Yang1,Yang2};
in statistical mechanics, it represents a sufficient condition such that the transfer matrix of a given statistical model commutes for different values of the spectral parameters \cite{Bax};
in integrable quantum field theory, the factorisation of the many-body scattering amplitudes into the product of pairwise scattering amplitudes is consistent if the YB equation is verified \cite{Zamolo}. 
Then, its importance grows with its fundamental role in the quantum inverse scattering method (or algebraic Bethe Ansatz) \cite{STF}. 
It also reveals to be important in the formulation of Hopf algebras and quantum groups, in knot theory or in the AdS-CFT correspondence.

A general construction providing matrix solutions of the YB equation in $\text{End}(V^{\otimes n})$ consists, roughly, in considering $V$ to be a representation of a 
quantum group\footnote{the solutions coming from the quantum group approach satisfies the non-braided YB equation; the braided version is simply obtained by multiplying by the permutation operator.} 
(for example, $U_q(sl_N)$), and to use the quasitriangular property of quantum groups (see \emph{e.g.} \cite{CP,Jim86}). This construction, very general, 
is technically intricate to apply in order to obtain explicit solutions (except for small ranks or small representations), since it involves to evaluate the (pseudo) universal $R$-matrix of 
affine quantum groups in representations (one can also replace the affine quantum group by the Yangian, a certain degeneration of it, to produce directly a certain ``scaling'' limit of 
the solutions associated to the affine quantum groups).

Having resulted in particular in the birth of quantum groups, the problem of constructing solutions of the YB equation has been and still is an active and fruitful field of investigation. Several methods have been developed and studied. One of these 
methods is the fusion procedure, allowing to build new solutions starting from a given one \cite{Jim86,KRS}. In principle, the fusion procedure allows to construct a solution of the 
YB equation acting on any representation which appears as a subrepresentation of a tensor product of fundamental representations of quantum groups. It is closely related to the Schur--Weyl 
duality by using projectors coming from the Hecke algebra (for type A) and BMW algebra (for other classical types), which are first obtained by a so-called fusion formula (see for example \cite{PdA}). 
In practice, when working in large representations, its algebraic technicality again makes it difficult to apply explicitly.

Among the other methods for constructing solutions of the YB equation, we mention the use of the spectral decomposition of 
the solutions, which aims at expressing them as sums of mutually orthogonal idempotents \cite{DGZ,Jim86}. Also, a 3D approach was developed recently 
and leads to explicit formulas for the matrix coefficients \cite{BM,Man14}. Finally, we mention also the recent calculation of some solutions for $sl_2$ from the study of 
the cohomology of certain algebraic varieties \cite{BZJ} and the direct computation providing a  classification of the $4\times 4$ $R$-matrices \cite{PR}.

In this paper, we focus on the Baxterisation method, and in a certain sense, we combine it with the fusion procedure.
This method (Baxterisation) has been proposed by V.F.R. Jones \cite{Jo}, in the framework of knot theory where solutions of the YB equation are obtained from representations 
of some algebra. It has been intensively used and numerous $R$-matrices have 
been obtained this way, see \emph{e.g.} \cite{ACDM,BM2000,CGX,Sn,BtB,IO,Jim86,KMN,YQL,ZGB}.
For an algebra $A$, a Baxterisation formula often refers to the following construction. Assume for simplicity that $A$ is generated by elements 
$\sigma_1,\dots,\sigma_{n-1}$ satisfying the braid relations. Then a Baxterisation formula in the algebra $A$ is an explicit formula for a function 
$\sigma_i(u)$ with values in the algebra $A$ (or more precisely, 
in the subalgebra generated by $\sigma_i$), such that the braided YB equation is satisfied inside $A$:
\[\sigma_i(u)\sigma_{i+1}(uv)\sigma_i(v)=\sigma_{i+1}(v)\sigma_i(uv)\sigma_{i+1}(u)\ .\]
Note that the expression of $\sigma_i(u)$ in terms of $\sigma_i$ should be the same for every $i$. Then consider a local representation of $A$, that is a representation of 
$A$ in $\text{End}(V^{\otimes n})$ for some vector space $V$ given in the form:
\[\sigma_i\mapsto \text{Id}^{\otimes i-1}\otimes \check R\otimes \text{Id}^{\otimes n-i-1}\ ,\]
where $\check R\in\text{End}(V\otimes V)$. Thus, applying the representation, the function $\sigma_i(u)$ is sent to a function with values in $\text{End}(V^{\otimes n})$ satisfying the 
usual braided YB equation for matrices. We emphasise that more general situation can be considered. For example the elements $\sigma_1,\dots,\sigma_n$ do not have to generate the whole 
algebra $A$, and the function $\sigma_i(u)$ could also involve other elements of the algebra $A$ than $\sigma_i$ (in this case, for a local representation one would 
require that the image of $\sigma_i(u)$ only acts on spaces number $i$ and $i+1$). 
Besides, one could start with elements of an algebra $\sigma_i$ satisfying different relations than the braid relation \cite{Sn,BtB}.

Probably the most classical example of a Baxterisation formula in an algebra is for the Hecke algebra. If $\sigma_1,\dots,\sigma_{n-1}$ denotes the standard generators of the
Hecke algebra $H_n(q)$, then a very simple formula provides a Baxterisation (see Formula (\ref{eq:Rhecke}) in this paper). There exists also a Baxterisation formula for the 
Birman--Murakami--Wenzl (BMW) algebra and for its degeneration the Brauer algebra (see for example \cite{IO}). The Brauer algebra is an example where additional generators 
(that is, in addition to the generators satisfying the braid relation) are needed and are present in the formula. This phenomenon also appears in the degeneration of the formula 
presented in this paper.

\paragraph{\textbf{Main result.}}
The Hecke algebra $H_n(q)$ admits a local representation on $\text{End}(V^{\otimes n})$ where $V$ is of dimension $N$ (for any $N>1$), 
and the solution of the braided YB equation obtained via the Baxterisation 
formula recovers the solution associated to the fundamental representation of the quantum group $U_q(sl_N)$.
This relation between the Hecke algebra and the fundamental representation $V$ of the quantum group $U_q(sl_N)$ can be understood thanks to the Schur--Weyl duality \cite{Jim86}.
In \cite{CPdA}, this duality has been extended for the representations $S^{(k)}_q(V)$ of $U_q(sl_N)$, the $q$-symmetrised $k$-th power of the fundamental ones.
In this case, the Hecke algebra must be replaced by new algebras, called fused Hecke algebra $H_{k,n}(q)$ in \cite{CPdA}. In the ``classical'' limit $q=1$, the fused Hecke algebra reduces to the algebra $H_{k,n}(1)$ called in \cite{CPdA}  the algebra of fused permutations. 

The main result of the present paper is a Baxterisation formula for these new algebras $H_{k,n}(q)$. 
We give an explicit formula providing solutions of the braided YB equation inside the fused Hecke algebra $H_{k,n}(q)$:
 \begin{equation*}
 \check R^{(k)}_i( u )=\sum_{p=0}^k (-q)^{k-p}\left[\begin{array}{c}k \\p\end{array}\right]^2_q\ 
 \frac{ (q^{-2}\, ;\, q^{-2})_{k-p}  }{ (uq^{-2p}\,;\, q^{-2})_{k-p}}\ \Sigma_i^{(k;p)}\ ,
 \end{equation*}
where the coefficients involve classical $q$-numbers (to be defined in the paper), and the elements $\Sigma_i^{(k;p)}$ are certain distinguished elements of $H_{k,n}(q)$, that we call (elementary) partial braidings. This formula generalises the formula for the Hecke algebra. Indeed the fused Hecke algebra $H_{k,n}(q)$ depends on an integer $k\geq 1$ and when $k=1$ it becomes 
simply the Hecke algebra $H_n(q)$ and our Baxterisation formula reduces to the usual formula in $H_n(q)$.

The classical or ``scaling'' limit $q=1$ of the above formula is trivial to perform and results immediately in a Baxterisation formula for the algebra of fused permutations $H_{k,n}(1)$ generalising 
the usual formula for the Yang solution in the symmetric group.

For $k\geq 1$, the algebra $H_{k,n}(q)$ admits a local representation on $S^{(k)}_q(V)$. The matrix solution of the braided YB equation obtained via the Baxterisation formula presented here recovers the solution associated to the representation $S^{(k)}_q(V)$ of the quantum group $U_q(sl_N)$. Thus we see the Baxterisation formula above as the first step in a two steps procedure to an explicit calculation of these matrix solutions, the second step consisting in sending this formula in the local representations of $H_{k,n}(q)$. This second step is a straightforward algebraic calculation of the action of the natural elements $\Sigma_i^{(k;p)}$, the elementary partial braiding, in the representations. In particular, we emphasise that the spectral parameters appears only in the coefficients above and not in the action of $\Sigma_i^{(k;p)}$. In particular, for $N=2$, we cover all finite-dimensional irreducible representations of $U_q(sl_2)$ (the integer $k\geq 1$ corresponds to the spin $k/2$).

One pleasant feature of the formula above is its uniformity for any $k\geq 1$ and the rather simple form of its coefficients. This is due to the choice of the elementary partial braidings in the expansions. Besides, for a given $k\geq 1$, this single formula produces an infinite number of matrix solutions (on spaces $S^{(k)}_q(V)$) since it covers all possible dimensions $N$ of $V$. In other words, the first step in the procedure evoked above is independent of the dimension $N$; this dimension appears only in the second step when applying the formula in the chosen local representation of $H_{k,n}(q)$.

In \cite{CPdA} the centralisers of the $U_q(sl_N)$-representations on tensor powers of $S^{(k)}_q(V)$ were obtained as quotients (depending on $N=\dim(V)$) of the fused Hecke algebra $H_{k,n}(q)$. Thus, the Baxterisation formula is also valid in all these centralisers. In the special case $N=2$, an expansion of the fused $R$-matrix for $sl_2$ has been computed previously in \cite{ZJ} in terms of some fused Temperley--Lieb generators (analogues of our partial elementary braidings in this case) and the question of an algebra underlying this construction was asked explicitly in \cite{Ma}. Applying our construction to this special case, a possible answer is the fused Hecke algebra $H_{k,n}(q)$ (and its quotient for $N=2$).

Finally, we note that one can consider also a mixed YB equation involving two different representations of $U_q(sl_N)$. In fact, we prove in this paper the generalisation of the formula above corresponding to the situation $S^{(k)}_q(V)\otimes S^{(l)}_q(V)$. As particular cases, this can be seen as a Baxterisation of other algebras introduced previously, such as the seam algebra \cite{MRR,LSA} and some
quotients of the Racah algebra introduced in \cite{CPV}.

\paragraph{\textbf{Organisation of the paper.}} In Section \ref{sec-fus-br}, we recall the definition of the fused Hecke algebra introduced in \cite{CPdA} 
and define the partial elementary braidings $\Sigma_i^{(k;p)}$. Section \ref{sec-YB} presents the main results of this paper: the Baxterisation formula for 
the fused Hecke algebra and the representations of this formula. The final section \ref{sec:pr} is devoted to the proof of the Baxterisation formula.

\paragraph{Acknowledgements.} Both authors are partially supported by Agence National de la Recherche Projet AHA
ANR-18-CE40-0001. N.Cramp\'e warmly thanks the university of Reims for hospitality
during his visit in the course of this investigation.

\section{Fused Hecke algebra}\label{sec-fus-br}

Let $n,k\in\mathbb{Z}_{>0}$\,. In what follows, the parameter $q$ is a non-zero complex number such that $1+q^2+\dots+q^{2(l-1)}\neq 0$ for $l=2,\dots,k$. 

In this section, we define the fused Hecke algebra $H_{k,n}(q)$ which is a particular case of the algebras introduced and studied in \cite{CPdA}. 
We refer to \cite{CPdA} for more details and more precision about the definition and the properties of $H_{k,n}(q)$.

\paragraph{Fused braids.} A fused braid is like a usual braid (more precisely, a braid diagram where two lines of $n$ dots are connected by strands) 
except that we have $k$ strands attached to each dot (so $k$ strands start from each dot in the top line and $k$ strands reach each dot in the bottom line). 
The total number of strands is then $nk$. To be more precise, the strands which are attached to the same dot are attached next to each other 
(so that we will often use the term ellipses instead of dots). We consider such diagrams up to homotopy, namely up to continuous moves of the strands leaving their end points fixed. 
Examples are drawn below.  

\paragraph{Examples.} $\bullet$ If $k=1$ then a fused braid is simply a usual braid.

$\bullet$ Here are examples of 6 fused braids when $n=3$ and $k=2$:
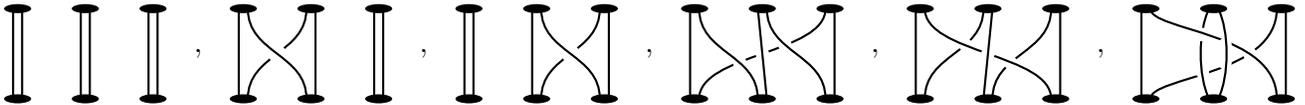
\begin{center}
\begin{tikzpicture}[scale=0.3]
\fill (1,2) ellipse (0.6cm and 0.2cm);\fill (1,-2) ellipse (0.6cm and 0.2cm);
\draw[thick] (0.8,2) -- (0.8,-2);\draw[thick] (1.2,2) -- (1.2,-2);
\fill (4,2) ellipse (0.6cm and 0.2cm);\fill (4,-2) ellipse (0.6cm and 0.2cm);
\draw[thick] (3.8,2) -- (3.8,-2);\draw[thick] (4.2,2) -- (4.2,-2);
\fill (7,2) ellipse (0.6cm and 0.2cm);\fill (7,-2) ellipse (0.6cm and 0.2cm);
\draw[thick] (6.8,2) -- (6.8,-2);\draw[thick] (7.2,2) -- (7.2,-2);
\node at (9,0) {$,$};
\fill (11,2) ellipse (0.6cm and 0.2cm);\fill (11,-2) ellipse (0.6cm and 0.2cm);!
\draw[thick] (10.8,2) -- (10.8,-2);\draw[thick] (13.8,2)..controls +(0,-2) and +(0,+2) .. (11.2,-2);\fill[white] (12.5,0) circle (0.4);
\fill (14,2) ellipse (0.6cm and 0.2cm);\fill (14,-2) ellipse (0.6cm and 0.2cm);
\draw[thick] (14.2,2) -- (14.2,-2);\draw[thick] (11.2,2)..controls +(0,-2) and +(0,+2) .. (13.8,-2);
\fill (17,2) ellipse (0.6cm and 0.2cm);\fill (17,-2) ellipse (0.6cm and 0.2cm);
\draw[thick] (16.8,2) -- (16.8,-2);\draw[thick] (17.2,2) -- (17.2,-2);
\node at (19,0) {$,$};
\fill (21,2) ellipse (0.6cm and 0.2cm);\fill (21,-2) ellipse (0.6cm and 0.2cm);
\draw[thick] (20.8,2) -- (20.8,-2);\draw[thick] (21.2,2) -- (21.2,-2);
\fill (24,2) ellipse (0.6cm and 0.2cm);\fill (24,-2) ellipse (0.6cm and 0.2cm);
\draw[thick] (23.8,2) -- (23.8,-2);\draw[thick] (26.8,2)..controls +(0,-2) and +(0,+2) .. (24.2,-2);\fill[white] (25.5,0) circle (0.4);
\fill (27,2) ellipse (0.6cm and 0.2cm);\fill (27,-2) ellipse (0.6cm and 0.2cm);
\draw[thick] (24.2,2)..controls +(0,-2) and +(0,+2) .. (26.8,-2);\draw[thick] (27.2,2) -- (27.2,-2);
\node at (29,0) {$,$};
\draw[thick] (36.8,2)..controls +(0,-2) and +(0,+2) .. (31.2,-2);\draw[thick] (37.2,2) -- (37.2,-2);
\fill[white] (33,-0.4) circle (0.3);\fill[white] (35,0.4) circle (0.3);;\fill[white] (34,0) circle (0.3);
\fill (31,2) ellipse (0.6cm and 0.2cm);\fill (31,-2) ellipse (0.6cm and 0.2cm);
\draw[thick] (30.8,2) -- (30.8,-2);\draw[thick] (31.2,2)..controls +(0,-2) and +(0,+2) .. (33.8,-2);
\fill (34,2) ellipse (0.6cm and 0.2cm);\fill (34,-2) ellipse (0.6cm and 0.2cm);
\draw[thick] (33.8,2) -- (34.2,-2);\draw[thick] (34.2,2)..controls +(0,-2) and +(0,+2) .. (36.8,-2);
\fill (37,2) ellipse (0.6cm and 0.2cm);\fill (37,-2) ellipse (0.6cm and 0.2cm);
\node at (39,0) {$,$};
\draw[thick] (46.8,2)..controls +(0,-2) and +(0,+2) .. (44.2,-2);
\draw[thick] (43.8,2)..controls +(0,-2) and +(0,+2) .. (41.2,-2);
\fill[white] (43,0.4) circle (0.3);\fill[white] (45,-0.4) circle (0.3);
\draw[thick] (41.2,2)..controls +(0,-2) and +(0,+2) .. (46.8,-2);
\fill[white] (44,0) circle (0.3);
\fill (41,2) ellipse (0.6cm and 0.2cm);\fill (41,-2) ellipse (0.6cm and 0.2cm);
\draw[thick] (40.8,2) -- (40.8,-2);
\fill (44,2) ellipse (0.6cm and 0.2cm);\fill (44,-2) ellipse (0.6cm and 0.2cm);
\draw[thick] (44.2,2) -- (43.8,-2);
\fill (47,2) ellipse (0.6cm and 0.2cm);\fill (47,-2) ellipse (0.6cm and 0.2cm);
\draw[thick] (47.2,2) -- (47.2,-2);
\node at (49,0) {$,$};
\draw[thick] (56.8,2)..controls +(0,-3) and +(0,+1) .. (51.2,-2);
\fill[white] (53.5,-0.8) circle (0.3);\fill[white] (54.5,-0.4) circle (0.3);
\draw[thick] (53.8,2)..controls +(-0.5,-1) and +(-0.5,1) .. (53.8,-2);
\fill[white] (53.5,0.85) circle (0.3);
\fill[white] (54,0.8) circle (0.3);\fill[white] (55.6,0) circle (0.3);
\draw[thick] (51.2,2)..controls +(0,-1) and +(0,+3) .. (56.8,-2);
\fill[white] (54.5,0.4) circle (0.3);
\draw[thick] (54.2,2)..controls +(0.5,-1) and +(0.5,1) .. (54.2,-2);
\draw[thick] (50.8,2) -- (50.8,-2);
\fill (51,2) ellipse (0.6cm and 0.2cm);\fill (51,-2) ellipse (0.6cm and 0.2cm);
\fill (54,2) ellipse (0.6cm and 0.2cm);\fill (54,-2) ellipse (0.6cm and 0.2cm);
\fill (57,2) ellipse (0.6cm and 0.2cm);\fill (57,-2) ellipse (0.6cm and 0.2cm);
\draw[thick] (57.2,2) -- (57.2,-2);
\end{tikzpicture}
\end{center}

Then we define a vector space as formal linear combination of fused braids with some local relations. 
The first relation in the following definition is the usual Hecke relation, valid locally for all crossings. 
The other relations are also local relations, valid for crossings near the ellipses.
\begin{defi}\label{vector-fused-braids}
 The $\CC$-vector space $H_{k,n}(q)$ is the vector space with basis indexed by fused braids quotiented by the following relations:
 \begin{itemize}
   \item[(i)] The Hecke relation:  
   \begin{center}
 \begin{tikzpicture}[scale=0.25]
\draw[thick] (0,2)..controls +(0,-2) and +(0,+2) .. (4,-2);
\fill[white] (2,0) circle (0.4);
\draw[thick] (4,2)..controls +(0,-2) and +(0,+2) .. (0,-2);
\node at (6,0) {$=$};
\draw[thick] (12,2)..controls +(0,-2) and +(0,+2) .. (8,-2);
\fill[white] (10,0) circle (0.4);
\draw[thick] (8,2)..controls +(0,-2) and +(0,+2) .. (12,-2);
\node at (17,0) {$-\,(q-q^{-1})$};
\draw[thick] (21,2) -- (21,-2);\draw[thick] (25,2) -- (25,-2);
\end{tikzpicture}
\end{center}
  \item[(ii)]  The idempotent relations: for top ellipses,
 \begin{center}
 \begin{tikzpicture}[scale=0.4]
\fill (2,2) ellipse (0.8cm and 0.2cm);
\draw[thick] (2.2,2)..controls +(0,-1.5) and +(1,1) .. (1.2,0);
\fill[white] (2,0.7) circle (0.2);
\draw[thick] (1.8,2)..controls +(0,-1.5) and +(-1,1) .. (2.8,0);
\node at (4.5,1) {$=$};
\node at (7,1) {$q$};
\fill (9,2) ellipse (0.8cm and 0.2cm);
\draw[thick] (8.8,2)..controls +(0,-1.5) and +(0.5,0.5) .. (8.2,0);
\draw[thick] (9.2,2)..controls +(0,-1.5) and +(-0.5,0.5) .. (9.8,0);

\node at (13,1) {and};

\fill (18,2) ellipse (0.8cm and 0.2cm);
\draw[thick] (17.8,2)..controls +(0,-1.5) and +(-1,1) .. (18.8,0);
\fill[white] (18,0.7) circle (0.2);
\draw[thick] (18.2,2)..controls +(0,-1.5) and +(1,1) .. (17.2,0);
\node at (20.5,1) {$= $};
\node at (23,1) {$q^{-1}$};
\fill (25,2) ellipse (0.8cm and 0.2cm);
\draw[thick] (24.8,2)..controls +(0,-1.5) and +(0.5,0.5) .. (24.2,0);
\draw[thick] (25.2,2)..controls +(0,-1.5) and +(-0.5,0.5) .. (25.8,0);
\end{tikzpicture}
\end{center}
and for bottom ellipses,
\begin{center}
 \begin{tikzpicture}[scale=0.4]
\fill (2,0) ellipse (0.8cm and 0.2cm);
\draw[thick] (1.8,0)..controls +(0,1.5) and +(-1,-1) .. (2.8,2);
\fill[white] (2,1.25) circle (0.2);
\draw[thick] (2.2,0)..controls +(0,1.5) and +(1,-1) .. (1.2,2);
\node at (4.5,1) {$=$};
\node at (7,1) {$q$};
\fill (9,0) ellipse (0.8cm and 0.2cm);
\draw[thick] (9.2,0)..controls +(0,1.5) and +(-0.5,-0.5) .. (9.8,2);
\draw[thick] (8.8,0)..controls +(0,1.5) and +(0.5,-0.5) .. (8.2,2);

\node at (13,1) {and};

\fill (18,0) ellipse (0.8cm and 0.2cm);
\draw[thick] (18.2,0)..controls +(0,1.5) and +(1,-1) .. (17.2,2);
\fill[white] (18,1.25) circle (0.2);
\draw[thick] (17.8,0)..controls +(0,1.5) and +(-1,-1) .. (18.8,2);
\node at (20.5,1) {$=$};
\node at (23,1) {$q^{-1}$};
\fill (25,0) ellipse (0.8cm and 0.2cm);
\draw[thick] (25.2,0)..controls +(0,1.5) and +(-0.5,-0.5) .. (25.8,2);
\draw[thick] (24.8,0)..controls +(0,1.5) and +(0.5,-0.5) .. (24.2,2);
\end{tikzpicture}
\end{center}
 \end{itemize}
\end{defi}
Near an ellipse, a ``positive'' crossing is replaced by $q$ while a ``negative'' crossing is replaced by $q^{-1}$ (this defines our convention for the words ``positive'' and ``negative''). 
Note that the Hecke relation near an ellipse is simply the sum of two idempotent relations. 

\paragraph{Usual Hecke algebra.} For $k=1$, the vector space $H_{1,n}(q)$ coincides with the vector space underlying the usual Hecke algebra $H_n(q)$. 
The multiplication in $H_{1,n}(q)$ is simply defined by concatenation of two diagrams and will be generalised below for $H_{k,n}(q)$.

We recall that the Hecke algebra $H_n(q)$ is generated by elements $\si_1,\dots,\si_{n-1}$ with defining relations:
\begin{equation}\label{rel-H}
\begin{array}{ll}
\si_i^2=(q-q^{-1})\si_i+1 & \text{for}\ i\in\{1,\dots,n-1\}\,,\\[0.2em]
\si_i\si_{i+1}\si_i=\si_{i+1}\si_i\si_{i+1}\ \ \  & \text{for $i\in\{1,\dots,n-2\}$}\,,\\[0.2em]
\si_i\si_j=\si_j\si_i\ \ \  & \text{for $i,j\in\{1,\dots,n-1\}$ such that $|i-j|>1$}\,.
\end{array}
\end{equation} 
Due to the braid relations (the second and third lines above), the Hecke algebra $H_n(q)$ is a quotient of the algebra of the braid group. 
As such, a braid diagram can be seen as an element of the algebra $H_n(q)$. The generators $\si_1,\dots,\si_{n-1}$ correspond 
to elementary positive crossing, namely $\si_i$ is the diagram where the $i$-th strand crosses over the $i+1$-th strand and all other strands are vertical.

We will need the notion of a standard diagram in $H_n(q)$. Let $w$ be a permutation in the symmetric group on $n$ elements. A standard diagram in $H_n(q)$ is a diagram connecting the dots in the top line with the dots in the bottom line according to the permutation $w$ with the following requirements: the number of crossings is minimal and all crossings are positive. For each $w$, we choose once and for all such a standard diagram and denote it $\si_w$.

\begin{rem}
The choice of a standard diagram for $w$ corresponds to a choice of a reduced expression $w$ in the symmetric group. Thanks to the braid relations in $H_n(q)$ and Matsumoto's lemma, any choice results in the same element $\si_w$ of $H_n(q)$. The set of elements $\si_w$ where $w$ runs over the symmetric group, forms the standard basis of $H_n(q)$.
\end{rem}

\paragraph{Multiplication of fused braids.} We define now a product on the vector space $H_{k,n}(q)$. For $L\in\mathbb{Z}_{\geq 0}$, we set:
$$\{L\}_q:=\frac{q^{2L}-1}{q^2-1}=1+q^2+\dots+q^{2(L-1)}\ .$$
Let $b,b'$ be two fused braids. We define the product $bb'$ in $H_{k,n}(q)$ as the result of the following procedure:
\begin{itemize}
\item \emph{(Concatenation)} We place the diagram of $b$ on top of the diagram of $b'$ by identifying the bottom ellipses of $b$ with the top ellipses of $b'$
\item \emph{(Removal of middle ellipses)} For a given ellipse in the middle row, there are $k$ strands arriving and $k$ strands leaving this ellipse. 
We remove this middle ellipse and replace it by the sum of all standard diagrams connecting these two sets of $k$ strands; in the sum, each diagram 
is pondered with the factor $q^\ell$ where $\ell$ is the number of crossings in the diagram, and the whole sum is divided by $\{2\}_q\{3\}_q\dots\{k\}_q$.
\end{itemize}
The removal of middle ellipses can be compactly expressed as follows. We replace locally each middle ellipse (using the $k$ strands arriving and the $k$ strands leaving) by the following element of the Hecke algebra $H_k(q)$:
\begin{equation}\label{symH}
\frac{1}{\{2\}_q\{3\}_q\dots\{k\}_q}\sum_w q^{\ell(w)}\si_w\ ,
\end{equation}
where the sum is over the permutation group on $k$ letters and $\ell(w)$ is the number of crossings in the standard diagram $\si_w$.

It turns out that this procedure (extended by linearity) provides $H_{k,n}(q)$ with a structure of an associative 
unital algebra, see \cite{CPdA}. This sums up the following facts. The procedure above results in a well-defined multiplication on $H_{k,n}(q)$: 
the associativity is immediate and the fused braid with only non-crossing vertical strands is the unit element for this multiplication. 
This follows from the fact that the element (\ref{symH}) is the $q$-symmetriser of $H_k(q)$, and as such, as we are going to recall in more details 
later in Section \ref{sec:pr}, is compatible with the idempotent relations and squares to itself (since the denominator in (\ref{symH}) 
coincides with the sum where $\si_w$ is evaluated to $q^{\ell(w)}$).

We call the resulting algebra \emph{the fused Hecke algebra} and we continue to use $H_{k,n}(q)$ to denote this algebra.

\begin{exa}\label{ex:rem} We illustrate the procedure to remove a middle ellipse when $k=2$:
 
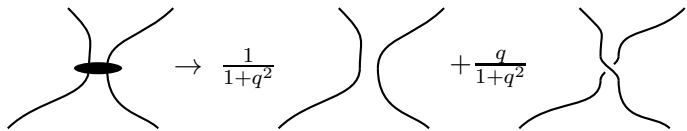
\begin{center}
 \begin{tikzpicture}[scale=0.4]
\fill (0,0) ellipse (0.8cm and 0.2cm);
\draw[thick] (-0.3,0)..controls +(0,1) and +(1,-1) .. (-1,2);
\draw[thick] (0.3,0)..controls +(0,1) and +(-1,-1) .. (2.5,2);
\draw[thick] (-0.3,0)..controls +(0,-1) and +(1,1) .. (-3,-2);
\draw[thick] (0.3,0)..controls +(0,-1.5) and +(-0.5,0.5) .. (2,-2);
\node at (3,0) {$\rightarrow$};
\node at (5,0) {$\frac{1}{1+q^2}$};
\draw[thick] (8.7,0)..controls +(0,1) and +(1,-1) .. (8,2);
\draw[thick] (9.3,0)..controls +(0,1) and +(-1,-1) .. (11.5,2);
\draw[thick] (8.7,0)..controls +(0,-1) and +(1,1) .. (6,-2);
\draw[thick] (9.3,0)..controls +(0,-1.5) and +(-0.5,0.5) .. (11,-2);
\node at (13,0) {$+\frac{q}{1+q^2}$};
\draw[thick] (16.7,0.4)..controls +(0,1) and +(1,-1) .. (16,2);
\draw[thick] (17.3,0.4)..controls +(0,1) and +(-1,-1) .. (19.5,2);
\draw[thick] (17.3,0.4)..controls +(0,-0.3) and +(0,0.3) .. (16.7,-0.4);
\fill[white] (17,0) circle (0.2);
\draw[thick] (16.7,0.4)..controls +(0,-0.3) and +(0,0.3) .. (17.3,-0.4);
\draw[thick] (16.7,-0.4)..controls +(0,-1) and +(1,1) .. (14,-2);
\draw[thick] (17.3,-0.4)..controls +(0,-1.5) and +(-0.5,0.5) .. (19,-2);
\end{tikzpicture}
\end{center}
\end{exa}

\begin{exa} Here is an example of a multiplication in $H_{2,2}(q)$ (using the Hecke and the idempotent relations in the second equality):
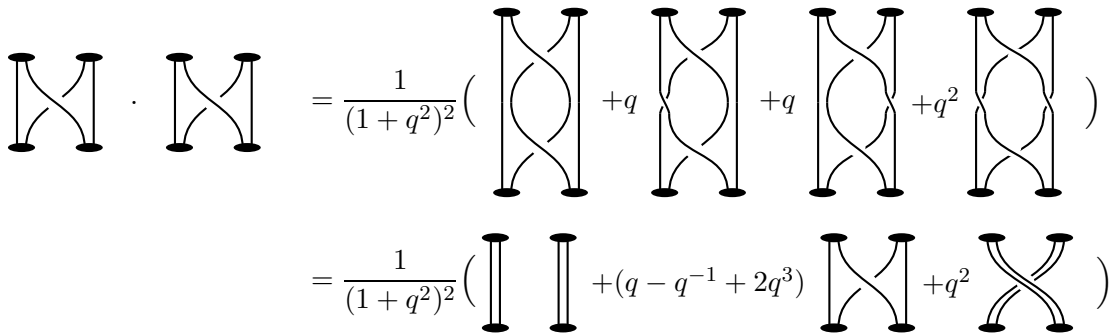
\begin{center}
\begin{tikzpicture}[scale=0.3]
\fill (1,2) ellipse (0.6cm and 0.2cm);\fill (1,-2) ellipse (0.6cm and 0.2cm);
\fill (4,2) ellipse (0.6cm and 0.2cm);\fill (4,-2) ellipse (0.6cm and 0.2cm);
\draw[thick] (3.8,2)..controls +(0,-2) and +(0,+2) .. (1.2,-2); \draw[thick] (4.2,2) -- (4.2,-2);
\fill[white] (2.5,0) circle (0.4);
\draw[thick] (0.8,2) -- (0.8,-2);\draw[thick] (1.2,2)..controls +(0,-2) and +(0,+2) .. (3.8,-2);  

\node at (6,0) {$.$};

\fill (8,2) ellipse (0.6cm and 0.2cm);\fill (8,-2) ellipse (0.6cm and 0.2cm);
\fill (11,2) ellipse (0.6cm and 0.2cm);\fill (11,-2) ellipse (0.6cm and 0.2cm);
\draw[thick] (10.8,2)..controls +(0,-2) and +(0,+2) .. (8.2,-2); \draw[thick] (11.2,2) -- (11.2,-2);
\fill[white] (9.5,0) circle (0.4);
\draw[thick] (7.8,2) -- (7.8,-2);\draw[thick] (8.2,2)..controls +(0,-2) and +(0,+2) .. (10.8,-2);  

\node at (17.5,0) {$=\displaystyle\frac{1}{(1+q^2)^2}\Bigl($};
\fill (22.5,4) ellipse (0.6cm and 0.2cm);
\fill (25.5,4) ellipse (0.6cm and 0.2cm);
\draw[thick] (25.3,4)..controls +(0,-2) and +(0,+2) .. (22.7,0); \draw[thick] (25.7,4) -- (25.7,0);
\fill[white] (24,2) circle (0.4);
\draw[thick] (22.3,4) -- (22.3,0);\draw[thick] (22.7,4)..controls +(0,-2) and +(0,+2) .. (25.3,0);  

\fill (22.5,-4) ellipse (0.6cm and 0.2cm);
\fill (25.5,-4) ellipse (0.6cm and 0.2cm);
\draw[thick] (25.3,0)..controls +(0,-2) and +(0,+2) .. (22.7,-4); \draw[thick] (25.7,0) -- (25.7,-4);
\fill[white] (24,-2) circle (0.4);
\draw[thick] (22.3,0) -- (22.3,-4);\draw[thick] (22.7,0)..controls +(0,-2) and +(0,+2) .. (25.3,-4);  

\node at (27.5,0) {$+ q$};
\fill (29.5,4) ellipse (0.6cm and 0.2cm);
\fill (32.5,4) ellipse (0.6cm and 0.2cm);
\draw[thick] (32.3,4)..controls +(0,-2) and +(0,+2) .. (29.7,0.5); \draw[thick] (32.7,4) -- (32.7,0);
\fill[white] (30.9,2) circle (0.4);
\draw[thick] (29.3,4) -- (29.3,0.5);\draw[thick] (29.7,4)..controls +(0,-2) and +(0,+2) .. (32.3,0);  

\draw[thick] (29.7,0.5)..controls +(0,-0.2) and +(0,+0.2) .. (29.3,-0.5);
\fill[white] (29.5,0) circle (0.2);
\draw[thick] (29.3,0.5)..controls +(0,-0.2) and +(0,+0.2) .. (29.7,-0.5);

\fill (29.5,-4) ellipse (0.6cm and 0.2cm);
\fill (32.5,-4) ellipse (0.6cm and 0.2cm);
\draw[thick] (32.3,0)..controls +(0,-2) and +(0,+2) .. (29.7,-4); \draw[thick] (32.7,0) -- (32.7,-4);
\fill[white] (30.8,-2.1) circle (0.4);
\draw[thick] (29.3,-0.5) -- (29.3,-4);\draw[thick] (29.7,-0.5)..controls +(0,-2) and +(0,+2) .. (32.3,-4);

\node at (34.5,0) {$+q$};
\fill (36.5,4) ellipse (0.6cm and 0.2cm);
\fill (39.5,4) ellipse (0.6cm and 0.2cm);
\draw[thick] (39.3,4)..controls +(0,-2) and +(0,+2) .. (36.7,0); \draw[thick] (39.7,4) -- (39.7,0.5);
\fill[white] (38.2,2.1) circle (0.4);
\draw[thick] (36.3,4) -- (36.3,0);\draw[thick] (36.7,4)..controls +(0,-2) and +(0,+2) .. (39.3,0.5);  

\draw[thick] (39.7,0.5)..controls +(0,-0.2) and +(0,+0.2) .. (39.3,-0.5);
\fill[white] (39.5,0) circle (0.2);
\draw[thick] (39.3,0.5)..controls +(0,-0.2) and +(0,+0.2) .. (39.7,-0.5);

\fill (36.5,-4) ellipse (0.6cm and 0.2cm);
\fill (39.5,-4) ellipse (0.6cm and 0.2cm);
\draw[thick] (39.3,-0.5)..controls +(0,-2) and +(0,+2) .. (36.7,-4); \draw[thick] (39.7,-0.5) -- (39.7,-4);
\fill[white] (38,-2) circle (0.4);
\draw[thick] (36.3,0) -- (36.3,-4);\draw[thick] (36.7,0)..controls +(0,-2) and +(0,+2) .. (39.3,-4);  

\node at (41.5,0) {$+q^2$};
\fill (43.5,4) ellipse (0.6cm and 0.2cm);
\fill (46.5,4) ellipse (0.6cm and 0.2cm);
\draw[thick] (46.3,4)..controls +(0,-2) and +(0,+2) .. (43.7,0.5); \draw[thick] (46.7,4) -- (46.7,0.5);
\fill[white] (45.1,2) circle (0.4);
\draw[thick] (43.3,4) -- (43.3,0.5);\draw[thick] (43.7,4)..controls +(0,-2) and +(0,+2) .. (46.3,0.5);  

\draw[thick] (43.7,0.5)..controls +(0,-0.2) and +(0,+0.2) .. (43.3,-0.5);
\fill[white] (43.5,0) circle (0.2);
\draw[thick] (43.3,0.5)..controls +(0,-0.2) and +(0,+0.2) .. (43.7,-0.5);

\draw[thick] (46.7,0.5)..controls +(0,-0.2) and +(0,+0.2) .. (46.3,-0.5);
\fill[white] (46.5,0) circle (0.2);
\draw[thick] (46.3,0.5)..controls +(0,-0.2) and +(0,+0.2) .. (46.7,-0.5);

\fill (43.5,-4) ellipse (0.6cm and 0.2cm);
\fill (46.5,-4) ellipse (0.6cm and 0.2cm);
\draw[thick] (46.3,-0.5)..controls +(0,-2) and +(0,+2) .. (43.7,-4); \draw[thick] (46.7,-0.5) -- (46.7,-4);
\fill[white] (44.9,-2) circle (0.4);
\draw[thick] (43.3,-0.5) -- (43.3,-4);\draw[thick] (43.7,-0.5)..controls +(0,-2) and +(0,+2) .. (46.3,-4);  

\node at (48.5,0) {$\Bigr)$};

\node at (17.5,-8) {$=\displaystyle\frac{1}{(1+q^2)^2}\Bigl($};
\fill (22,-6) ellipse (0.6cm and 0.2cm);\fill (22,-10) ellipse (0.6cm and 0.2cm);
\draw[thick] (21.8,-6) -- (21.8,-10);\draw[thick] (22.2,-6) -- (22.2,-10);
\fill (25,-6) ellipse (0.6cm and 0.2cm);\fill (25,-10) ellipse (0.6cm and 0.2cm);
\draw[thick] (24.8,-6) -- (24.8,-10);\draw[thick] (25.2,-6) -- (25.2,-10);

\node at (31,-8) {$+(q-q^{-1}+2q^3)$};

\fill (37,-6) ellipse (0.6cm and 0.2cm);\fill (37,-10) ellipse (0.6cm and 0.2cm);
\fill (40,-6) ellipse (0.6cm and 0.2cm);\fill (40,-10) ellipse (0.6cm and 0.2cm);
\draw[thick] (39.8,-6)..controls +(0,-2) and +(0,+2) .. (37.2,-10);  \draw[thick] (40.2,-6) -- (40.2,-10);
\fill[white] (38.5,-8) circle (0.4);
\draw[thick] (36.8,-6) -- (36.8,-10);\draw[thick] (37.2,-6)..controls +(0,-2) and +(0,+2) .. (39.8,-10);  

\node at (42,-8) {$+q^2$};

\fill (44,-6) ellipse (0.6cm and 0.2cm);\fill (44,-10) ellipse (0.6cm and 0.2cm);
\fill (47,-6) ellipse (0.6cm and 0.2cm);\fill (47,-10) ellipse (0.6cm and 0.2cm);
\draw[thick] (46.8,-6)..controls +(0,-2) and +(0,+2) .. (43.8,-10);  
\draw[thick] (47.2,-6)..controls +(0,-2) and +(0,+2) .. (44.2,-10);  
\fill[white] (45.5,-8) circle (0.4);
\draw[thick] (43.8,-6)..controls +(0,-2) and +(0,+2) .. (46.8,-10);  
\draw[thick] (44.2,-6)..controls +(0,-2) and +(0,+2) .. (47.2,-10);  

\node at (49,-8) {$\Bigr)$};

\end{tikzpicture}
\end{center}
\end{exa}

\begin{rem}
The fused Hecke algebra is well-defined if $q^2=1$. In this case, we denote it $H_{k,n}(1)$ and we call it the algebra of fused permutations 
(recall that the Hecke algebra for $q^2=1$ is isomorphic to the group algebra of the symmetric group). 
We emphasise that the definition of $H_{k,n}(1)$ is entirely combinatorial and its multiplication is much easier to perform than in the general case. 
The algebra $H_{k,n}(q)$ is a deformation of $H_{k,n}(1)$ (see \cite{CPdA}). 
\end{rem}

\paragraph{Partial elementary braidings.}
In the fused Hecke algebras $H_{k,n}(q)$, there are elements called partial elementary braidings which are of particular interest for the following.
\begin{defi} Let $i \in\{1,\dots,n-1\}$ and $0\leq p\leq k$. The partial elementary braiding $\Sigma_i^{(k;p)}\in H_{k,n}(q)$ corresponds to the diagram in which the $p$ 
rightmost strands starting from ellipse $i$ pass over the $p$ leftmost strands starting from ellipse $i+1$. All other strands are vertical. 
\end{defi}

For any $i$, the element $\Sigma_i^{(k;0)}$ is the identity element of $H_{k,n}(q)$ and $\Sigma_i^{(k;k)}$ is called the $i$-th full elementary braiding 
(or simply $i$-th elementary braiding). It is immediate to see that the full elementary braidings $\Sigma_i^{(k;k)}$ satisfy the braid relations
\begin{equation}
 \begin{array}{cl}
 \Sigma_i^{(k;k)}\Sigma_{i+1}^{(k;k)}\Sigma_i^{(k;k)}=\Sigma_{i+1}^{(k;k)}\Sigma_i^{(k;k)}\Sigma_{i+1}^{(k;k)}\ \  & \\[0.5em]
 \Sigma_i^{(k;k)}\Sigma_{j}^{(k;k)}=\Sigma_{j}^{(k;k)}\Sigma_i^{(k;k)} & \text{if $|i-j|>1$.}
 \end{array}
\end{equation}
Thus these elements $\Sigma_1^{(k;k)},\dots,\Sigma_{n-1}^{(k;k)}$ generate a subalgebra of $H_{k,n}(q)$ which is a quotient of the algebra of the braid group. Note that the partial elementary braidings do not satisfy the first braid relations. They satisfy in general $\Sigma_i^{(k;p)}\Sigma_{j}^{(k;p')}=\Sigma_{j}^{(k;p')}\Sigma_i^{(k;p)}$ if $|i-j|>1$, for any $p,p'$. 

In addition, the full elementary braiding has, for any $i$, a minimal polynomial of degree $k+1$ (see \cite{CPdA}):
\begin{equation}
 \prod_{l=0}^k\Bigl(\Sigma^{(k;k)}_i-(-1)^{k+l} q^{-k+l(l+1)}\Bigr)=0\ .
\end{equation}
For $q$ generic in the sense that it is not a root of the unity, the elements $\Sigma_i^{(k;p)}$ for $p=0,\dots,k-1$ are generated by $\Sigma_i^{(k;k)}$. In particular, if $n=2$, the algebra $H_{k,2}(q)$ is generated by $\Sigma_1^{(k;k)}$. Note that in general the algebra $H_{k,n}(q)$ is not generated by $\Sigma_1^{(k;k)},\dots,\Sigma_{n-1}^{(k;k)}$ (in fact it happens only if $n=2$ or if $k>1$).

If $q^2=1$, the situation is drastically different. In addition to the braid relation, the elementary braidings $\Sigma_1^{(k;k)},\dots,\Sigma_{n-1}^{(k;k)}$ are involutions and thus generate a subalgebra in $H_{n,k}(1)$ isomorphic to the algebra of the symmetric group. The elements $\Sigma_i^{(k;p)}$ for $p=1,\dots,k-1$ are not obtained in this subalgebra.

\begin{exa} $\bullet$ We recall that $H_{1,n}(q)$ is isomorphic to the Hecke algebra  $H_{n}(q)$ and we have the following identification $\Sigma_i^{(1;1)}=\sigma_i$.\\
 $\bullet$ For $k=2$, the partial elementary braidings are:
\begin{center}
 \begin{tikzpicture}[scale=0.23]
\node at (-3,0) {$\Sigma_i^{(2;0)}:=$};
\node at (2,3) {$1$};\fill (2,2) ellipse (0.6cm and 0.2cm);\fill (2,-2) ellipse (0.6cm and 0.2cm);
\draw[thick] (1.8,2) -- (1.8,-2);\draw[thick] (2.2,2) -- (2.2,-2);
\node at (4,0) {$\dots$};
\draw[thick] (5.8,2) -- (5.8,-2);\draw[thick] (6.2,2) -- (6.2,-2);
\node at (6,3) {$i-1$};\fill (6,2) ellipse (0.6cm and 0.2cm);\fill (6,-2) ellipse (0.6cm and 0.2cm);
\node at (10,3) {$i$};\fill (10,2) ellipse (0.6cm and 0.2cm);\fill (10,-2) ellipse (0.6cm and 0.2cm);
\node at (14,3) {$i+1$};\fill (14,2) ellipse (0.6cm and 0.2cm);\fill (14,-2) ellipse (0.6cm and 0.2cm);

\draw[thick] (9.8,2)..controls +(0,-2) and +(0,+2) .. (9.8,-2);
\draw[thick] (10.2,2)..controls +(0,-2) and +(0,+2) .. (10.2,-2);
\fill[white] (12,0) circle (0.5);
\draw[thick] (14.2,2)..controls +(0,-2) and +(0,+2) .. (14.2,-2);
\draw[thick] (13.8,2)..controls +(0,-2) and +(0,+2) .. (13.8,-2);

\draw[thick] (17.8,2) -- (17.8,-2);\draw[thick] (18.2,2) -- (18.2,-2);
\node at (18,3) {$i+2$};\fill (18,2) ellipse (0.6cm and 0.2cm);\fill (18,-2) ellipse (0.6cm and 0.2cm);
\node at (20,0) {$\dots$};
\draw[thick] (21.8,2) -- (21.8,-2);\draw[thick] (22.2,2) -- (22.2,-2);\fill (22,2) ellipse (0.6cm and 0.2cm);\fill (22,-2) ellipse (0.6cm and 0.2cm);
\node at (22,3) {$n$};
\end{tikzpicture}
\end{center}

\begin{center}
 \begin{tikzpicture}[scale=0.23]
\node at (-3,0) {$\Sigma_i^{(2;1)}:=$};
\node at (2,3) {$1$};\fill (2,2) ellipse (0.6cm and 0.2cm);\fill (2,-2) ellipse (0.6cm and 0.2cm);
\draw[thick] (1.8,2) -- (1.8,-2);\draw[thick] (2.2,2) -- (2.2,-2);
\node at (4,0) {$\dots$};
\draw[thick] (5.8,2) -- (5.8,-2);\draw[thick] (6.2,2) -- (6.2,-2);
\node at (6,3) {$i-1$};\fill (6,2) ellipse (0.6cm and 0.2cm);\fill (6,-2) ellipse (0.6cm and 0.2cm);
\node at (10,3) {$i$};\fill (10,2) ellipse (0.6cm and 0.2cm);\fill (10,-2) ellipse (0.6cm and 0.2cm);
\node at (14,3) {$i+1$};\fill (14,2) ellipse (0.6cm and 0.2cm);\fill (14,-2) ellipse (0.6cm and 0.2cm);

\draw[thick] (9.8,2) -- (9.8,-2);
\draw[thick] (13.8,2)..controls +(0,-2) and +(0,+2) .. (10.2,-2);
\fill[white] (12,0) circle (0.4);
\draw[thick] (10.2,2)..controls +(0,-2) and +(0,+2) .. (13.8,-2);
\draw[thick] (14.2,2) -- (14.2,-2);

\draw[thick] (17.8,2) -- (17.8,-2);\draw[thick] (18.2,2) -- (18.2,-2);
\node at (18,3) {$i+2$};\fill (18,2) ellipse (0.6cm and 0.2cm);\fill (18,-2) ellipse (0.6cm and 0.2cm);
\node at (20,0) {$\dots$};
\draw[thick] (21.8,2) -- (21.8,-2);\draw[thick] (22.2,2) -- (22.2,-2);\fill (22,2) ellipse (0.6cm and 0.2cm);\fill (22,-2) ellipse (0.6cm and 0.2cm);
\node at (22,3) {$n$};
\end{tikzpicture}
\end{center}

\begin{center}
 \begin{tikzpicture}[scale=0.23]
\node at (-3,0) {$\Sigma_i^{(2;2)}:=$};
\node at (2,3) {$1$};\fill (2,2) ellipse (0.6cm and 0.2cm);\fill (2,-2) ellipse (0.6cm and 0.2cm);
\draw[thick] (1.8,2) -- (1.8,-2);\draw[thick] (2.2,2) -- (2.2,-2);
\node at (4,0) {$\dots$};
\draw[thick] (5.8,2) -- (5.8,-2);\draw[thick] (6.2,2) -- (6.2,-2);
\node at (6,3) {$i-1$};\fill (6,2) ellipse (0.6cm and 0.2cm);\fill (6,-2) ellipse (0.6cm and 0.2cm);
\node at (10,3) {$i$};\fill (10,2) ellipse (0.6cm and 0.2cm);\fill (10,-2) ellipse (0.6cm and 0.2cm);
\node at (14,3) {$i+1$};\fill (14,2) ellipse (0.6cm and 0.2cm);\fill (14,-2) ellipse (0.6cm and 0.2cm);

\draw[thick] (13.8,2)..controls +(0,-2) and +(0,+2) .. (9.8,-2);
\draw[thick] (14.2,2)..controls +(0,-2) and +(0,+2) .. (10.2,-2);
\fill[white] (12,0) circle (0.5);
\draw[thick] (10.2,2)..controls +(0,-2) and +(0,+2) .. (14.2,-2);
\draw[thick] (9.8,2)..controls +(0,-2) and +(0,+2) .. (13.8,-2);

\draw[thick] (17.8,2) -- (17.8,-2);\draw[thick] (18.2,2) -- (18.2,-2);
\node at (18,3) {$i+2$};\fill (18,2) ellipse (0.6cm and 0.2cm);\fill (18,-2) ellipse (0.6cm and 0.2cm);
\node at (20,0) {$\dots$};
\draw[thick] (21.8,2) -- (21.8,-2);\draw[thick] (22.2,2) -- (22.2,-2);\fill (22,2) ellipse (0.6cm and 0.2cm);\fill (22,-2) ellipse (0.6cm and 0.2cm);
\node at (22,3) {$n$};
\end{tikzpicture}
\end{center}
\end{exa}

\section{Baxterisation formula in $H_{k,n}(q)$ \label{sec-YB}}

For $L\in\mathbb{Z}_{\geq 0}$, we define the $q$-numbers by $[L]_q:=\frac{q^L-q^{-L}}{q-q^{-1}}$, the $q$-factorial by $[L]_q!:=[1]_q[2]_q\dots[L]_q$, the $q$-binomials by
 \begin{equation}
  \left[\begin{array}{c}L \\p\end{array}\right]_q:=\frac{[L]_q!}{[L-p]_q![p]_q!}\ ,
 \end{equation}
and the $q$-Pochhammer symbol by
\begin{equation}
 (a\, ;\, q)_p=\prod_{r=0}^{p-1}(1-aq^r)\ .
\end{equation}
By convention, we have $[0]_q!=\left[\begin{array}{c}L \\0\end{array}\right]_q=(a\, ;\, q)_0=1$. 

\begin{thm}\label{th:bax}
The following elements in $H_{k,n}(q)$, for $1\leq i\leq n-1$:
 \begin{equation}
 \check R^{(k)}_i( u )=\sum_{p=0}^k (-q)^{k-p}\left[\begin{array}{c}k \\p\end{array}\right]^2_q\ 
 \frac{ (q^{-2}\, ;\, q^{-2})_{k-p}  }{ (uq^{-2p}\,;\, q^{-2})_{k-p}}\ \Sigma_i^{(k;p)}\ , \label{eq:R}
 \end{equation}
satisfy the braided Yang--Baxter equations, for $1\leq i\leq n-2$,
\begin{equation}
\check  R^{(k)}_i(u)\check R^{(k)}_{i+1}(uv)\check R^{(k)}_i(v)=\check R^{(k)}_{i+1}(v) \check R^{(k)}_i(uv) \check R^{(k)}_{i+1}(v)\ .
\end{equation}
\end{thm}
The proof of this theorem is postponed to Section \ref{sec:pr}.
The parameter $u$ in relation \eqref{eq:R} is usually called spectral parameter and $\check R^{(k)}( u )$ is called spectral dependent braided $R$-matrix or simply $R$-matrix. We see this theorem as a Baxterisation formula, in the sense recalled in the introduction, for the fused Hecke algebra $H_{k,n}(q)$.

We emphasize the importance of the choice of the partial elementary braidings for the expansion of the $R$-matrix. 
Indeed, this choice results in a rather simple and elegant form for the coefficients. 
This would not have been so if we had chosen for example the powers of the full elementary braiding $\Sigma_i^{(k;k)}$ for the expansion (see also the remarks following the formula for the scaling limit).

\begin{exa}$\bullet$ Let $k=1$. The previous formula \eqref{eq:R} reduces to
\begin{equation}
  \check R_i(u):=\check R^{(1)}_i(u)=\sigma_i  - \frac{q-q^{-1}}{1-u} \ , \label{eq:Rhecke}
\end{equation}
where we have used the isomorphism $H_{1,n}(q)\sim H_n(q)$. It is well-known that this R-matrix satisfy the Yang--Baxter equation and it can be checked by a quick calculation.\\
$\bullet$ Let $k=2$. The baxterised $R$-matrix becomes
\begin{equation}\label{exaR(2)}
\check R^{(2)}_i(u)=\Sigma_i^{(2;2)} - \frac{(q+q^{-1})(q^2-q^{-2}) }{1-uq^{-2}}\Sigma_i^{(2;1)} +q^2 \frac{(1-q^{-2})(1-q^{-4})}{(1-u)(1-uq^{-2})} \Sigma_i^{(2;0)} \ .
\end{equation}
Note that a direct verification that the Yang--Baxter equation is satisfied in $H_{2,n}(q)$ is already a very difficult task. We will follow a different approach, in the spirit of the fusion procedure, for the proof of Theorem \ref{th:bax}.
\end{exa}

\paragraph{Scaling limit.} Let $q=\exp(h/2)$ and consider the limit $h\rightarrow 0$ in the previous $R$-matrix \eqref{eq:R} where we rescale also the spectral parameter as follows $u=\exp(h\mu)$.
This limit behaves well and gives 
\begin{equation}
 \check R^{(k)}_i( \mu )=\sum_{p=0}^k \left(\begin{array}{c}k \\p\end{array}\right)^2\ 
 \frac{ (k-p)!  }{ (\mu-p)(\mu-p-1)\dots (\mu-k+1)   }\ \Sigma_i^{(k;p)}\ . \label{eq:Ra}
 \end{equation}
Note that this solution lives in the algebra $H_{k,n}(1)$ of fused permutations. The elements $\Sigma_i^{(k;p)}$ in this algebra are represented by similar diagram but with no crossing (that is, the positive and negative crossings are identified). These $R$-matrices given by \eqref{eq:Ra} verify an additive version of the braided Yang--Baxter equation, for $1\leq i\leq n-2$,
\begin{equation}
\check  R^{(k)}_i(\mu)\check R^{(k)}_{i+1}(\mu+\nu)\check R^{(k)}_i(\nu)=\check R^{(k)}_{i+1}(\nu) \check R^{(k)}_i(\mu+\nu) \check R^{(k)}_{i+1}(\nu)\ .
\end{equation}
Note that the limit here is direct to perform, thanks to the convenient choice of the partial elementary braidings for the expansion of the $R$-matrix. For another choice of generators (such as the powers of the full braiding element) this limit should be more involved, since in the situation $q^2=1$, the powers of the full braiding element do not generate the partial braidings.

\paragraph{Representations.} One motivation for considering the algebra $H_{k,n}(q)$ is that it admits local representations which will allow to obtain from (\ref{eq:R}) genuine 
matrix solutions of the Yang--Baxter equation.

Let $N>1$ and $V$ a vector space of dimension $N$. Denote $W=S^{(k)}_q(V)$ the quantum $k$-th symmetric power of $V$. The subspace $W$ is the image of the $q$-symmetriser on $k$ spaces 
applied on $V^{\otimes k}$. Denote by $S_{[1,k]}$ the $q$-symmetriser. An explicit formula for the $q$-symmetriser inside the Hecke algebra $H_k(q)$ was recalled in (\ref{symH}) 
and a different formula is recalled in the next Section in (\ref{def-P}). The natural action of $H_k(q)$ on $V^{\otimes k}$ is also recalled below.

We found in \cite{CPdA} that the fused Hecke algebra $H_{k,n}(q)$ admits a local representation on the space
\[W\otimes W\otimes \dots \otimes W\ \ \ \ \text{($n$ factors)}\,,\]
where by local, we mean that the elementary partial braidings $\Sigma_i^{(k;p)}$ act non-trivially only on the factors $i$ and $i+1$. So as an immediate corollary of the preceding theorem, we obtain the following solutions of the Yang--Baxter equation.
\begin{coro}
The images of the $R$-matrices $\check R^{(k)}_i( u )$ in the representation of $H_{k,n}(q)$ on $W^{\otimes n}$ are solutions of the Yang--Baxter equation in $\text{End}(W^{\otimes n})$.
\end{coro}
\begin{rem}
Note that, for a given $k\geq 1$, the algebra $H_{k,n}(q)$ provides an infinite family of matrix solutions since the dimension $N$ of $V$ can be arbitrary. From the Schur--Weyl duality, the subspace $W$ is an irreducible representation of the quantum group $U_q(sl_N)$ and so the formula (\ref{eq:R}) can be seen as a formula for expressing the corresponding solution of the Yang--Baxter equation. For $N=2$, the solution corresponds to the spin $k/2$ representation of $U_q(sl_2)$.
\end{rem}
One use of Formula (\ref{eq:R}) is the following. In order to calculate explicitly the matrix solution, one only needs to calculate the images of the partial braidings $\Sigma_i^{(k;p)}$ in $\text{End}(W^{\otimes n})$, and then it remains only to apply (\ref{eq:R}). We give more details. Of course, it is enough to consider the situation $n=2$, since the representation is local, and so we can forget the index $i$ in $\Sigma_i^{(k;p)}$

Let $(e_1,\dots,e_N)$ be a basis of $V$. First let us recall the usual action of the Hecke algebra $H_m(q)$ on the space $V^{\otimes m}$. The generator $\si_i$ goes to $\text{Id}^{\otimes i-1}\otimes \check R\otimes \text{Id}^{\otimes m-1-i}$, where $\check R$ is the standard Hecke $R$-matrix acting on $V\otimes V$:
\[\check{R}(e_i\otimes e_j):=\left\{\begin{array}{ll}
q\,e_i\otimes e_j\ \  & \text{if $i=j$,}\\[0.8em]
e_j\otimes e_i+(q-q^{-1})\,e_i\otimes e_j & \text{if $i<j$,}\\[0.4em]
e_j\otimes e_i & \text{if $i>j$.}
\end{array}\right.\ \ \ \text{where $i,j=1,\dots,N$.}\]

Now we are ready to give the action of the partial elementary braidings $\Sigma^{(k;p)}$ of $H_{k,2}(q)$ on $W^{\otimes 2}$. From \cite{CPdA} (and see also (\ref{eq:Si}) later), we have that the action of $\Sigma^{(k;p)}$ on $W\otimes W$ is obtained through the action on $V^{\otimes k}\otimes V^{\otimes k}=V^{\otimes 2k}$ of the following element:
\[S_{[1,k]}S_{[k+1,2k]}  (\check{R}_k \check{R}_{k+1}\dots \check{R}_{k+p-1}) (\check{R}_{k-1} \check{R}_{k}\dots \check{R}_{k+p-2})  \dots  (\check{R}_{k-p+1} \check{R}_{k-p+1}\dots \check{R}_{k})  S_{[1,k]}S_{[k+1,2k]} \ .
\]
In this formula, $S_{[1,k]}$ is the $q$-symmetriser on the first $k$ spaces while $S_{[k+1,2k]}$ is the $q$-symmetriser on the last $k$ spaces. Note that due to the presence of this projector $S_{[1,k]}S_{[k+1,2k]}$ on both sides, this action indeed restricts to an action on $W\otimes W$. The formula between the two projectors is best visualised with a diagram like the one just after Formula (\ref{eq:Si}).

So at this point it becomes a straightforward (technically complicated) computational task to calculate the action of $\Sigma^{(k;p)}$ on $W\otimes W$. We note that one can use as a basis of the vector space $W$:
\[\{S_{[1,k]}\cdot(e_{i_1}\otimes e_{i_2}\otimes\dots\otimes e_{i_k})\}_{1\leq i_1\leq i_2\leq\dots\leq i_k\leq N}\ .\]
The dimension of $W$ is $\left(\begin{array}{c}k+N-1\\k\end{array}\right)$.

\begin{rem}
If $q^2=1$ then $S_{[1,k]}$ is the usual symmetriser on $k$ arguments (the sum of all permutations in the symmetric group divided by $k!$), and the element between the two projectors in the formula for the action of $\Sigma^{(k;p)}$ is a simple permutation operator, an involution, exchanging the block of letters $k-p+1,\dots,k$ with the block of letters $k+1,\dots,k+p$ (see the diagram in (\ref{eq:Si})). So in the situation $q^2=1$, the action of  $\Sigma^{(k;p)}$ is rather easy to calculate (one can also take the limit $q\to 1$ in the general formula, as in the example below, but it is much simpler to calculate it directly in the situation $q^2=1$).
\end{rem}

\begin{exa}
Let $k=2$ and $N=2$. We set $\lambda=q-q^{-1}$ and $2_q=q+q^{-1}$. A basis of $W$ is $(v_1,v_2,v_3)$ with $v_1=e_1\otimes e_1$, $v_2=\frac{1}{2_q}(qe_1\otimes e_2+e_1\otimes e_2)$ and $v_3=e_2\otimes e_2$. We use the natural basis $(v_i\otimes v_j)_{i,j=1,2,3}$ of $W\otimes W$ ordered lexicographically. In this basis, the action of the elements $\Sigma_1^{(2;1)},\Sigma_1^{(2;2)}$ is given, respectively, by
\[\left(\begin{array}{ccccccccc}
\!q\! & . & . & . & . & . & . & . & .\\
. & \!\!\!q-\frac{q^{-2}}{2_q}\!\!\! & . & \frac{1}{2_q} & . & . & . & . & .\\
. & . & \lambda & . & \frac{q^2}{(2_q)^2} & . & . & . & .\\
. & \frac{1}{2_q} & . & \frac{1}{2_q} & . & . & . & . & .\\
. & . & 1 & . & \!\!\!\frac{2-q^{-2}}{2_q}\!\!\! & . & \!\!q^{-2}\!\! & . & .\\
. & . & . & . & . & \!\!\!q-\frac{q^{-2}}{2_q}\!\!\! & . & \frac{1}{2_q} & .\\
. & . & . & . & \frac{1}{(2_q)^2} & . & . & . & .\\
. & . & . & . & . & \frac{1}{2_q} & . & \frac{1}{2_q} & .\\
. & . & . & . & . & . & . & . & \!q\!
\end{array}\right)\,,\ \ \left(\begin{array}{ccccccccc}
\!q^4\! & . & . & . & . & . & . & . & .\\
. & \!\!q^22_q\lambda\!\! & . & q^2 & . & . & . & . & .\\
. & . & \!\!q2_q\lambda^2\!\! & . & q^2\lambda & . & 1 & . & .\\
. & q^2 & . & . & . & . & . & . & .\\
. & . & \!\!\!(2_q)^2\lambda\!\!\! & . & q^2 & . & . & . & .\\
. & . & . & . & . & \!\!q^22_q\lambda\!\! & . & q^2 & .\\
. & . & 1 & . & . & . & . & . & .\\
. & . & . & . & . & q^2 & . & . & .\\
. & . & . & . & . & . & . & . & \!q^4\!
\end{array}\right)\ . \]
From this, the formula (\ref{exaR(2)}) gives directly the solution of the Yang--Baxter equation on $W\otimes W$.
\end{exa}

\section{Proof of the Baxterisation formula \label{sec:pr}}

This section is devoted to the proof of Theorem \ref{th:bax}. The main steps of the proof are the following:
\begin{itemize}
 \item the fused Hecke algebra $H_{k,n}(q)$ is isomorphic to a certain projection of the usual Hecke algebra $H_{nk}(q)$;
 \item we write the elementary partial braidings in terms of elements in $H_{nk}(q)$;
 \item we show that a certain product of the usual $R$-matrices \eqref{eq:Rhecke}  of the Hecke algebra $H_{nk}(q)$ can be expanded as a linear combination of the elementary partial braidings;
 \item we identify this sum with the r.h.s. of \eqref{eq:R} and use the factorised form in terms of usual $R$-matrices to show that it satisfies the braided Yang--Baxter equation (this is where the ideas of the fusion procedure appear).
\end{itemize}

\paragraph{Projected Hecke algebra.}

In the Hecke algebra $H_{nk}(q)$, we consider the elements, for $1\leq i < j \leq nk-1$,
\begin{equation}\label{def-P}
S_{[i,j]}:=\frac{1}{[j-i+1 ]_q!} \ \prod_{i\leq a\leq j-1}^{\longrightarrow}\check R_a(q^{2(a-i+1)})\ \check R_{a-1}(q^{2(a-i)})\ \dots \check R_i(q^2) \ ,
\end{equation} 
where $\check R_a(u)$ is given by Formula \eqref{eq:Rhecke} and $\displaystyle \prod_{i\leq a\leq j-1}^{\longrightarrow}$ means that the product is ordered from left to right when the index $a$ increases. This element is the image of $S_{[1,j-i+1]}$ through the embedding of $H_{j-i+1}(q)$ in $H_{nk}(q)$ given by $\sigma_a\mapsto\sigma_{a+i-1}$.

The element $S_{[1,j-i+1]}$ is the so-called $q$-symmetriser of the Hecke algebra $H_{j-i+1}(q)$. In fact, Formula (\ref{def-P}) is the simplest case of the fusion formula for 
the Hecke algebra expressing a complete set of primitive idempotents as products of $R$-matrices \cite{IMO}. The $q$-symmetrisers are well-known 
objects (see \cite{IO09,Jim86}). We recall below the main properties that we will use in the following.
A formula alternative to (\ref{def-P}) is the following sum:
\[S_{[i,j]}=\frac{q^{-\frac{(j-i+1)(j-i)}{2}}}{[j-i+1]_q!}\sum_w q^{\ell(w)}\si_w\ ,\]
where the sum runs over the set of permutations of the letters $\{i,i+1,\dots,j\}$, the elements $\si_w$ are the corresponding standard basis elements of $H_{nk}(q)$, and $\ell(w)$ is the number of crossings in the standard diagram $\si_w$ (equivalently, the length of $w$).
The element $S_{[i,j]}$ is a partial $q$-symmetriser satisfying in particular
\begin{equation}\label{rel-sym}
S_{[i,j]}^2=S_{[i,j]}\ \ \ \ \ \text{and}\ \ \ \ \ \ \si_a S_{[i,j]}=S_{[i,j]}\si_a=q S_{[i,j]}\,,\ a=i,\dots,j-1. 
\end{equation}
This implies that $S_{[i,j]}=S_{[i,j]}S_{[i',j']}$ if $i\leq i'<j'\leq j$. Finally, a recursion formula for these partial $q$-symmetrisers is:
\begin{equation}\label{rec-sym}
S_{[i,j+1]}=\frac{1}{[j-i]_q}\sum_{a=i}^{j+1}q^{i-a}\si_a\si_{a+1}\dots\si_j S_{[i,j]}\ .
\end{equation}

Inside the Hecke algebra $H_{nk}(q)$, we consider also the element
\begin{equation}\label{def-Pkn}
P^{(k)}:=S_{[1,k]}S_{[k+1,2k]}\dots  S_{[(n-1)k+1,nk]}\ .
\end{equation}
This is an idempotent (each factor is itself an idempotent and commutes with the others) and it allows us to construct $P^{(k)} H_{nk}(q)P^{(k)}$ which is an algebra with the unit $P^{(k)}$. Here we call this algebra the projected Hecke algebra. 
In \cite{CPdA}, the following proposition is established. 
\begin{prop}\label{prop-PHP}
 The fused Hecke algebra $H_{k,n}(q)$ is isomorphic to the projected Hecke algebra $P^{(k)}  H_{nk}(q) P^{(k)}$.
\end{prop}
Thanks to this proposition, we can transfer the proof of Theorem \ref{th:bax} from $H_{k,n}(q)$ to $P^{(k)} H_{nk}(q)P^{(k)}$.

\paragraph{Elementary partial braidings in the projected Hecke algebra.}
In this paragraph, we restrict ourselves to the case $n=2$. This case is sufficient when we want to study only one $R$-matrix.
By using the isomorphism between $P^{(k)} H_{2k}(q)P^{(k)}$ and $H_{k,2}(q)$, the partial elementary braiding $\Sigma^{(k;p)}$ reads as follows
\begin{equation}\label{eq:Si}
 \Sigma^{(k;p)}= P^{(k)}  (\sigma_k \sigma_{k+1}\dots \sigma_{k+p-1}) (\sigma_{k-1} \sigma_{k}\dots \sigma_{k+p-2})  \dots  (\sigma_{k-p+1} \sigma_{k-p+1}\dots \sigma_{k})  P^{(k)} \ ,
\end{equation}
where $P^{(k)}=S_{[1,k]}S_{[k+1,2k]}$. Recall that $p\in\{0,\dots,k\}$ (when $p=0$, the element is simply $P^{(k)}$). The formula for $\Sigma^{(k;p)}$ is best visualised on the following diagram:
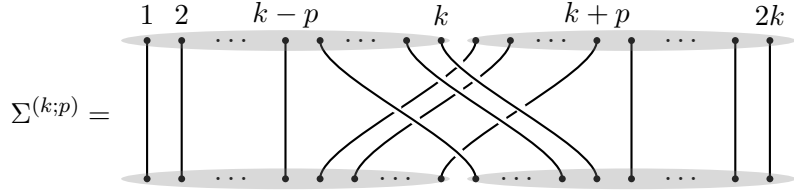
\begin{center}
 \begin{tikzpicture}[scale=0.23]
\node at (-3,0) {$\Sigma^{(k;p)}=$};
\node at (2,5.5) {$1$};\fill (2,4) circle (0.2cm);\fill (2,-4) circle (0.2cm);\draw[thick] (2,4) -- (2,-4);
\node at (4,5.5) {$2$};\fill (4,4) circle (0.2cm);\fill (4,-4) circle (0.2cm);\draw[thick] (4,4) -- (4,-4);
\node at (7,4) {$\dots$};\node at (7,-4) {$\dots$};
\node at (10,5.5) {$k-p$};\fill (10,4) circle (0.2cm);\fill (10,-4) circle (0.2cm);\draw[thick] (10,4) -- (10,-4);
\fill (12,4) circle (0.2cm);\fill (12,-4) circle (0.2cm);
\node at (14.5,4) {$\dots$};\node at (16.5,-4) {$\dots$};
\fill (17,4) circle (0.2cm);\fill (14,-4) circle (0.2cm);
\node at (19,5.5) {$k$};\fill (19,4) circle (0.2cm);\fill (19,-4) circle (0.2cm);

\fill (21,4) circle (0.2cm);\fill (21,-4) circle (0.2cm);
\fill (23,4) circle (0.2cm);\fill (26,-4) circle (0.2cm);
\node at (25.5,4) {$\dots$};\node at (23.5,-4) {$\dots$};
\node at (28,5.5) {$k+p$};\fill (28,4) circle (0.2cm);\fill (28,-4) circle (0.2cm);
\fill (30,4) circle (0.2cm);\fill (30,-4) circle (0.2cm);\draw[thick] (30,4) -- (30,-4);
\node at (33,4) {$\dots$};\node at (33,-4) {$\dots$};
\fill (36,4) circle (0.2cm);\fill (36,-4) circle (0.2cm);\draw[thick] (36,4) -- (36,-4);
\node at (38,5.5) {$2k$};\fill (38,4) circle (0.2cm);\fill (38,-4) circle (0.2cm);\draw[thick] (38,4) -- (38,-4);

\draw[thick] (21,4)..controls +(0,-2) and +(0,+2) .. (12,-4);
\draw[thick] (23,4)..controls +(0,-2) and +(0,+2) .. (14,-4);
\draw[thick] (28,4)..controls +(0,-2) and +(0,+2) .. (19,-4);

\fill[white] (20,2.7) circle (0.3);

\fill[white] (20,1.1) circle (0.3);
\fill[white] (20,-2.5) circle (0.3);
\fill[white] (21,1.8) circle (0.3);\fill[white] (19,1.8) circle (0.3);
\fill[white] (16.4,0.1) circle (0.3);\fill[white] (17.5,-0.6) circle (0.3);
\fill[white] (23.6,0.1) circle (0.3);\fill[white] (22.5,-0.6) circle (0.3);

\draw[thick] (19,4)..controls +(0,-2) and +(0,+2) .. (28,-4);
\draw[thick] (17,4)..controls +(0,-2) and +(0,+2) .. (26,-4);
\draw[thick] (12,4)..controls +(0,-2) and +(0,+2) .. (21,-4);

\fill[fill=gray,opacity=0.3] (10,4) ellipse (9.5cm and 0.6cm);
\fill[fill=gray,opacity=0.3] (10,-4) ellipse (9.5cm and 0.6cm);
\fill[fill=gray,opacity=0.3] (30,4) ellipse (9.5cm and 0.6cm);
\fill[fill=gray,opacity=0.3] (30,-4) ellipse (9.5cm and 0.6cm);

\end{tikzpicture}
\end{center}
The 4 shaded ellipses surrounding the dots $1,2, \dots k$ and $k+1,\dots,2k$ in the previous diagram represent the $q$-symmetrisers $S_{[1,k]}$ and $S_{[k+1,2k]}$.

More generally, let $l\geq k$ and define the following elements in $H_{k+\ell}(q)$, for $0\leq p\leq k$,
\begin{equation}
 \Sigma^{(k,\ell;p)}= P^{(k,\ell)}  (\sigma_k \sigma_{k+1}\dots \sigma_{\ell+p-1}) (\sigma_{k-1} \sigma_{k}\dots \sigma_{\ell+p-2})  \dots  (\sigma_{k-p+1} \sigma_{k-p+2}\dots \sigma_{\ell})  P^{(\ell,k)} \ ,
\end{equation}
where $P^{(k,\ell)}=S_{[1,k]}S_{[k+1,k+\ell]}$ and $P^{(\ell,k)}=S_{[1,\ell]}S_{[\ell+1,k+\ell]}$.
We get $\Sigma^{(k,k;p)}=\Sigma^{(k;p)}$.
These elements are visualised on the following diagrams:
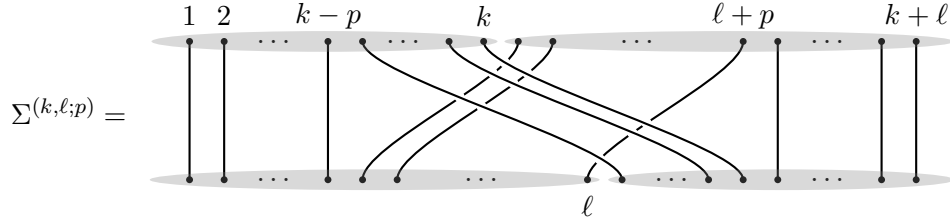
\begin{center}
 \begin{tikzpicture}[scale=0.23]
\node at (-5,0) {$\Sigma^{(k,\ell;p)}=$};
\node at (2,5.5) {$1$};\fill (2,4) circle (0.2cm);\fill (2,-4) circle (0.2cm);\draw[thick] (2,4) -- (2,-4);
\node at (4,5.5) {$2$};\fill (4,4) circle (0.2cm);\fill (4,-4) circle (0.2cm);\draw[thick] (4,4) -- (4,-4);
\node at (7,4) {$\dots$};\node at (7,-4) {$\dots$};
\node at (10,5.5) {$k-p$};\fill (10,4) circle (0.2cm);\fill (10,-4) circle (0.2cm);\draw[thick] (10,4) -- (10,-4);
\fill (12,4) circle (0.2cm);\fill (12,-4) circle (0.2cm);
\node at (14.5,4) {$\dots$};\node at (19,-4) {$\dots$};
\fill (17,4) circle (0.2cm);\fill (14,-4) circle (0.2cm);
\node at (19,5.5) {$k$};\fill (19,4) circle (0.2cm);\fill (32,-4) circle (0.2cm);

\fill (21,4) circle (0.2cm);\fill (25,-4) circle (0.2cm);
\fill (23,4) circle (0.2cm);\fill (27,-4) circle (0.2cm);
\node at (28,4) {$\dots$};\node at (30,-4) {$\dots$};
\node at (34,5.5) {$\ell+p$};\fill (34,4) circle (0.2cm);\fill (34,-4) circle (0.2cm);
\fill (36,4) circle (0.2cm);\fill (36,-4) circle (0.2cm);\draw[thick] (36,4) -- (36,-4);
\node at (39,4) {$\dots$};\node at (39,-4) {$\dots$};
\fill (42,4) circle (0.2cm);\fill (42,-4) circle (0.2cm);\draw[thick] (42,4) -- (42,-4);
\node at (44,5.5) {$k+\ell$};\fill (44,4) circle (0.2cm);\fill (44,-4) circle (0.2cm);\draw[thick] (44,4) -- (44,-4);

\node at (25,-5.5) {$\ell$};

\draw[thick] (21,4)..controls +(0,-2) and +(0,+2) .. (12,-4);
\draw[thick] (23,4)..controls +(0,-2) and +(0,+2) .. (14,-4);
\draw[thick] (34,4)..controls +(0,-2) and +(0,+2) .. (25,-4);

\fill[white] (20.2,2.8) circle (0.3);
\fill[white] (20.7,1.6) circle (0.3);
\fill[white] (25.7,-2.8) circle (0.3);
\fill[white] (21.5,2.2) circle (0.3);\fill[white] (19.5,2.2) circle (0.3);
\fill[white] (17.6,0.8) circle (0.3);\fill[white] (18.9,0.2) circle (0.3);
\fill[white] (28.4,-0.7) circle (0.3);\fill[white] (27.6,-1.3) circle (0.3);

\draw[thick] (19,4)..controls +(0,-2) and +(0,+2) .. (34,-4);
\draw[thick] (17,4)..controls +(0,-2) and +(0,+2) .. (32,-4);
\draw[thick] (12,4)..controls +(0,-2) and +(0,+2) .. (27,-4);

\fill[fill=gray,opacity=0.3] (9.8,4) ellipse (10cm and 0.6cm);
\fill[fill=gray,opacity=0.3] (12.7,-4) ellipse (13cm and 0.6cm);
\fill[fill=gray,opacity=0.3] (33.2,4) ellipse (13cm and 0.6cm);
\fill[fill=gray,opacity=0.3] (36.2,-4) ellipse (10cm and 0.6cm);

\end{tikzpicture}
\end{center}

\paragraph{Product of $R$-matrices.}
Let us now define, in $H_{k+\ell}(q) $,
\begin{equation}
 \check R^{(k,\ell)}(u) := P^{(k,\ell)} \prod_{1\leq a\leq k}^{\longleftarrow} \check R_a\left(uq^{2(1-a)}\right)\check R_{a+1}\left(uq^{2(2-a)}\right) \dots \check R_{a+\ell-1}\left(uq^{2(\ell-a)}\right)   P^{(\ell,k)}\ ,
 \label{eq:defkl}
\end{equation}
where  $\check R_a\left(u\right)$ is given by \eqref{eq:Rhecke}. The arrow means that the product is ordered from right to left when the index $a$ increases. It is a straightforward calculation, using repeatedly the braided Yang--Baxter equation satisfied by $\check R_i(u)$, to show that:
\begin{equation}\label{commPR}
P^{(k,\ell)}\check R^{(k,\ell)}(u)=\check R^{(k,\ell)}(u)P^{(\ell,k)}\ .
\end{equation}
As a consequence, only one idempotent (any one of the two) is actually needed in (\ref{eq:defkl}). We refer to Remark \ref{rem-mixed} for more information on the element $\check R^{(k,\ell)}$.

\begin{lem}\label{lem:te}
Let $1\leq k\leq \ell$. The element $\check R^{(k,\ell)}(u)$ can be rewritten as follows
 \begin{equation}
  \check R^{(k,\ell)}(u)=\sum_{p=0}^{k}  a^{(k,\ell)}_p(u)\   \Sigma^{(k,\ell;p)}\ ,      \label{eq:lem}
 \end{equation}
 where
  \begin{equation}
   a^{(k,\ell)}_p(u) = (-q)^{k-p}  \frac{ (q^{-2}\, ;\, q^{-2})_{k-p}  }{ (uq^{-2p}\,;\, q^{-2})_{k-p}}\left[\begin{array}{c}k \\p\end{array}\right]_q
   \left[\begin{array}{c}\ell \\k-p\end{array}\right]_q\ .
 \end{equation}
\end{lem}
\proof We prove this lemma by induction w.r.t. $k$ and $\ell$. For $k=\ell=1$, the definition \eqref{eq:defkl} becomes
 \begin{equation}
 \check R^{(1,1)}(u) := \check R_1\left(u\right)=\sigma_1  - \frac{q-q^{-1}}{1-u} =\Sigma_1^{(1,1;1)} - \frac{q-q^{-1}}{1-u}\Sigma_1^{(1,1;0)}
\end{equation}
which proves the Lemma for $k=\ell=1$. 
For $k=1$ and $\ell> 1$, the definition \eqref{eq:defkl} becomes
\begin{eqnarray}
 \check R^{(1,\ell)}(u) := S_{[2,\ell+1]}\check  R_1\left(u\right) \dots \check R_\ell(uq^{2(\ell-1)}) S_{[1,\ell]}\ .
\end{eqnarray}
We use $S_{[2,\ell+1]}=S_{[2,\ell+1]} S_{[2,\ell]}$ and the consequence of (\ref{commPR}) to obtain:
\begin{eqnarray}
 \check R^{(1,\ell)}(u) = S_{[2,\ell+1]}\check R_{[1,\ell]}^{(1,\ell-1)}(u) \check R_\ell(uq^{2(\ell-1)}) S_{[1,\ell]}\ .
\end{eqnarray}
The index of $\check R_{[1,\ell]}^{(1,\ell-1)}(u)$ indicates that we use the embedding of $H_{\ell}(q)$ in $H_{\ell+1}(q)$ given by $\sigma_i \mapsto \sigma_i$.
If we suppose that Formula (\ref{eq:lem}) holds for $\ell-1$, then 
\begin{eqnarray}
 \check R^{(1,\ell)}(u) = S_{[2,\ell+1]}\left(  \Sigma^{(1,\ell-1;1)}   +a^{(1,\ell-1)}_0(u)\   \Sigma^{(1,\ell-1;0)}  \right)\left(\sigma_\ell  - \frac{q-q^{-1}}{1-uq^{2(\ell-1)}} \right)   S_{[1,\ell]}\ .
\end{eqnarray}
From the definition of $\Sigma^{(1,\ell-1;1)}$, one obtains
\begin{eqnarray}
 \check R^{(1,\ell)}(u) &=& S_{[2,\ell+1]}\left(  \sigma_1\dots \sigma_\ell -\frac{q-q^{-1}}{1-uq^{2(\ell-1)}}  \sigma_1\dots \sigma_{\ell-1}
 +a^{(1,\ell-1)}_0(u)\sigma_\ell  -\frac{q-q^{-1}}{1-uq^{2(\ell-1)}} a^{(1,\ell-1)}_0(u) \right)   S_{[1,\ell]}\nonumber \\
 &=& \Sigma^{(1,\ell;1)} +S_{[2,\ell+1]}\left(  -\frac{q-q^{-1}}{1-uq^{2(\ell-1)}}  q^{\ell-1}
 +a^{(1,\ell-1)}_0(u)q  -\frac{q-q^{-1}}{1-uq^{2(\ell-1)}} a^{(1,\ell-1)}_0(u) \right)   S_{[1,\ell]}\ .\label{eq:aa}
\end{eqnarray}
To obtain the previous equality, we have used  $S_{[2,\ell+1]}\sigma_\ell=q S_{[2,\ell+1]}$ and $ \sigma_j S_{[1,\ell]}=q S_{[1,\ell]}$ (for $j=1,2, \dots ,\ell-1$).
By noticing that the expression in parenthesis in \eqref{eq:aa} is equal to $a^{(1,\ell)}_0(u)$, we prove by induction the Lemma for $k=1$ and $\ell> 1$.

Let $1<k \leq \ell$. 
By definition, one gets in $H_{k+\ell}(q)$
\begin{eqnarray}
 \check R^{(k,\ell)}(u) &=&S_{[1,k]}S_{[k+1,k+\ell]}\ \  \check R_k(uq^{2(1-k)})\check R_{k+1}(uq^{2(2-k)} )\dots \check R_{k+\ell-1} (uq^{2(\ell-k)})\nonumber  \\
 &&\qquad \qquad \qquad \times \qquad\qquad \vdots \nonumber \\ 
 &&\qquad \qquad \qquad \times \check R_2(uq^{-2})\check R_3(u)\dots \check R_{\ell+1} (uq^{2(\ell-2)})\nonumber  \\
 &&\qquad \qquad \qquad \times \check R_1(u)\check R_2(uq^2)\dots \check R_\ell (uq^{2(\ell-1)}) \ \  S_{[1,\ell]}S_{[\ell+1,k+\ell]}  \ .
 \end{eqnarray}
We use that $S_{[1,k]}=S_{[1,k]}S_{[2,k]}$ and the consequence of (\ref{commPR}) to obtain
 \begin{eqnarray}
 \check R^{(k,\ell)}(u)  &=&S_{[1,k]} \  \check R_{[2,k+\ell]}^{(k-1,\ell)}(uq^{-2})\  \check R_{[1,\ell+1]}^{(1,\ell)}(u) \  S_{[\ell+1,k+\ell]}
 \end{eqnarray} 
 where the index of $\check R_{[2,k+\ell]}^{(k-1,\ell)}(u)$ indicates that we use the embedding of $H_{k+\ell-1}(q)$ into  $H_{k+\ell}(q)$ given by $\sigma_i \mapsto \sigma_{i+1}$, while the index of $\check R_{[1,\ell+1]}^{(1,\ell)}(u) $ indicates that we use the embedding of $H_{\ell+1}(q)$ into  $H_{k+\ell}(q)$ given by $\sigma_i \mapsto \sigma_{i}$.

We use Relation (\ref{eq:lem}) for $\check R^{(k-1,\ell)}(u)$ and the previous result for $\check R^{(1,\ell)}(u)$ to get
 \begin{eqnarray}
  \check R^{(k,\ell)}(u)  &=&S_{[1,k]} \ \sum_{p=0}^{k-1} a_p^{(k-1,\ell)}(uq^{-2}) \Sigma_{[2,k+\ell]}^{(k-1,\ell;p)} \ 
  \Big(\sigma_1\dots \sigma_\ell+a_0^{(1,\ell)}(u)\Big) S_{[1,\ell]}  S_{[\ell+1,k+\ell]}\ .
 \end{eqnarray}
Let $p\in\{0,\dots,k-1\}$. We replace $\Sigma_{[2,k+\ell]}^{(k-1,\ell;p)}$ using its definition. First we have:
 \begin{eqnarray}
 && S_{[1,k]} \Sigma_{[2,k+\ell]}^{(k-1,\ell;p)}\sigma_1 \dots \sigma_\ell S_{[1,\ell]}  S_{[\ell+1,k+\ell]} \nonumber \\
  &=&S_{[1,k]}S_{[k+1,k+\ell]}   (\sigma_k \dots \sigma_{\ell+p}) \dots( \sigma_{k-p+1} \dots \sigma_{\ell+1})  S_{[2,\ell+1]}  \sigma_1\dots \sigma_\ell S_{[1,\ell]}  S_{[\ell+1,k+\ell]}\nonumber\\
  &=&S_{[1,k]}S_{[k+1,k+\ell]}   (\sigma_k \dots \sigma_{\ell+p}) \dots( \sigma_{k-p+1} \dots \sigma_{\ell+1})   (\sigma_1\dots \sigma_\ell)  S_{[1,\ell]}  S_{[\ell+1,k+\ell]}\nonumber\\
  &=&q^{k-p-1}  \Sigma^{(k,\ell;p+1)}\ .
 \end{eqnarray}
In the second equality, we use $S_{[2,\ell+1]}  \sigma_1\dots \sigma_\ell=\sigma_1\dots \sigma_\ell S_{[1,\ell]}$ (this is (\ref{commPR}) when $u\to\infty$); in the last, we notice that $\si_1,\dots,\si_{k-p-1}$ in the last parenthesis commute with all elements on their left, and give a factor $q$ when multiplied to $S_{[1,k]}$ (this also works for $p=0$ since there is nothing on their left then except $S_{[1,k]}S_{[k+1,k+\ell]}$).

Second, we have, using (\ref{rec-sym}) for $S_{[2,\ell+1]}$ and $S_{[2,\ell]}S_{[1,\ell]}=S_{[1,\ell]}$,
\begin{eqnarray}
&&  S_{[1,k]} \Sigma_{[2,k+\ell]}^{(k-1,\ell;p)}S_{[1,\ell]}   S_{[\ell+1,k+\ell]} \label{calc-lem1} \\
  &=&S_{[1,k]}S_{[k+1,k+\ell]}   (\sigma_k \dots \sigma_{\ell+p}) \dots( \sigma_{k-p+1} \dots \sigma_{\ell+1})  S_{[2,\ell+1]} S_{[1,\ell]}  S_{[\ell+1,k+\ell]}\nonumber\\
  &=&\frac{1}{[\ell]_q} \sum_{a=2}^{\ell+1}q^{2-a} S_{[1,k]}S_{[k+1,k+\ell]}  \underbrace{(\sigma_k \dots \sigma_{\ell+p}) \dots( \sigma_{k-p+1} \dots \sigma_{\ell+1})}_{=:x}  (\sigma_a\sigma_{a+1}\dots \sigma_\ell)  S_{[1,\ell]}  S_{[\ell+1,k+\ell]}\nonumber\ .
  \end{eqnarray} 
One checks that $x$ commutes with $\si_2,\dots,\si_{k-p-1}$, while $x\si_b=\si_{b+p}x$ if $b>k-p$. Thus we have
\[S_{[1,k]}S_{[k+1,\ell+k]}\,x\si_b\ =\ q\,S_{[1,k]}S_{[k+1,\ell+k]}\,x\  \ \ \ \ \ \text{for any $n\neq k-p$.}\]
Moreover, moving the rightmost element in each parenthesised factor of $x$ and multiplying it to $S_{[\ell+1,k+\ell]}$, one finds that
\[x\,S_{[\ell+1,k+\ell]}=q^p(\sigma_k \dots \sigma_{\ell+p-1}) \dots( \sigma_{k-p+1} \dots \sigma_{\ell})\,S_{[\ell+1,k+\ell]}\ .\]
So the element whose calculation was started in (\ref{calc-lem1}) is equal to
\begin{eqnarray}
  &=&\frac{1}{[\ell]_q} \sum_{a=2}^{k-p}q^{k-p+2-2a} S_{[1,k]}S_{[k+1,k+\ell]}   (\sigma_k \dots \sigma_{\ell+p}) \dots( \sigma_{k-p+1} \dots \sigma_{\ell+1}) (\sigma_{k-p}\sigma_{k-p+1}\dots \sigma_\ell)  S_{[1,\ell]}  S_{[\ell+1,k+\ell]}\nonumber\\
  &&+\frac{1}{[\ell]_q} \sum_{a=k-p+1}^{\ell+1}q^{\ell+p+3-2a} S_{[1,k]}S_{[k+1,k+\ell]}   (\sigma_k \dots \sigma_{\ell+p-1}) \dots( \sigma_{k-p+1} \dots \sigma_{\ell})   S_{[1,\ell]}  S_{[\ell+1,k+\ell]}\nonumber\\
  &=& \frac{[k-p-1]_q}{[\ell]_q} \Sigma^{(k,\ell;p+1)}  +q^{2p-k+1} \frac{[\ell+p-k+1]_q}{[\ell]_q} \Sigma^{(k,\ell;p)}\ .
 \end{eqnarray}
A straightforward verification shows that
 \begin{eqnarray}
&&  q^{-k+1}\frac{[\ell-k+1]_q}{[\ell]_q} a_0^{(k-1,\ell)}(uq^{-2})a_0^{(1,\ell)}(u)=a_0^{(k,\ell)}(u)\,,\\ 
&&a_{p-1}^{(k-1,\ell)}(uq^{-2})\left( q^{k-p}+\frac{[k-p]_q}{[\ell]_q}a_0^{(1,\ell)}(u)\right)+a_{p}^{(k-1,\ell)}(uq^{-2})q^{2p-k+1} \frac{[\ell+p-k+1]_q}{[\ell]_q}a_0^{(1,\ell)}(u)=a_p^{(k,\ell)}(u)\,,\qquad\\
&&a_{k-1}^{(k-1,\ell)}(uq^2)=a_{k}^{(k,\ell)}(u)\ .
 \end{eqnarray}
So we have proved that Relation \eqref{eq:lem} is also valid for $k$ and $\ell$, which proves the lemma by induction.\endproof 

\paragraph{Braided Yang--Baxter equation.} We are now in position to prove Theorem \ref{th:bax}. 
By using the Lemma \ref{lem:te}, we see that in $P^{(k)}H_{kn}(q)P^{(k)}$, one gets, for $i=1,\dots n-1$,
\begin{equation}
 \check R_i^{(k)}(u) = \check R^{(k,k)}_i(u)
\end{equation}
where the index $i$ of $\check R^{(k,k)}_i(u)$ indicates that we use the embedding of $H_{2k}(q)$ into $H_{nk}(q)$ given by $\sigma_a \mapsto \sigma_{a+(i-1)k}$.
Therefore, we can use the factorised form given by \eqref{eq:defkl} for $\check R_i^{(k)}(u)$. Namely, one gets
\begin{equation}
 \check R_i^{(k)}(u) := P^{(k)} \prod_{1\leq a\leq k}^{\longleftarrow} \check R_{a+(i-1)k}\left(uq^{2(1-a)}\right)\check R_{a+1+(i-1)k}\left(uq^{2(2-a)}\right) \dots \check R_{a+\ell-1+(i-1)k}\left(uq^{2(\ell-a)}\right)   P^{(k)}\ ,
 \label{eq:defkk}
\end{equation}
Relation (\ref{commPR}) gives in this case that $\check R_i^{(k)}(u)$ commutes with the projector $P^{(k)}$ so that
\begin{equation}
 \check R_i^{(k)}(u) = P^{(k)} \prod_{1\leq a\leq k}^{\longleftarrow} \check R_{a+(i-1)k}\left(uq^{2(a-1)}\right) \check R_{a+1+(i-1)k}\left(uq^{2(a-2)}\right) \dots \check R_{a-1+ik}\left(uq^{2(a-k)}\right)  \ . \label{eq:Rf}
\end{equation}
It is then a standard computation to show that $\check R_i^{(k)}(u)$ given by \eqref{eq:Rf} satisfies the braided Yang--Baxter equation (see remark below). This concludes the proof of Theorem \ref{th:bax}.

\begin{rem}\label{rem-mixed}
We recall that in fact the elements $\check R^{(k,l)}(u)$ given in (\ref{eq:defkl}) satisfy a braided Yang--Baxter equation in general (we used this when $k=l$; for $k\neq l$ one can call it a mixed Yang--Baxter equation). Namely, for $k,l,m\geq 1$ and $n=k+l+m$, we have the following equality in $H_{n}(q)$: 
\[\check R^{(k,l)}(u)\check R^{(k,m)}_{[l+1,n]}(uv)\check R^{(l,m)}(v)=\check R^{(l,m)}_{[k+1,n]}(v)\check R^{(k,m)}(uv)\check R^{(k,l)}_{[m+1,n]}(u)\ ,\]
where an index $[a+1,n]$ indicates that we use the embedding of $H_{n-a}(q)$ in $H_{n}(q)$ given by $\si_i\mapsto\si_{i+a}$. This is a standard and straightforward computation in the context of the fusion procedure (along the lines of some calculations detailed for example in \cite{PdA}). One has to use repeatedly the braided Yang--Baxter equation satisfied by $\check R_i(u)$. The first step is to show that:
\[S_{[1,a]}\check R^{(a,b)}(u)=\check R^{(a,b)}(u)S_{[b+1,a+b]}\ \ \ \ \text{and}\ \ \ \ S_{[a+1,a+b]}\check R^{(a,b)}(u)=\check R^{(a,b)}(u)S_{[1,b]}\ .\]
Using this, in order to check the braided Yang--Baxter equation in $H_n(q)$, one can first move all projectors on one side, and then it remains only to check the braided Yang--Baxter equation satisfied by the product without the projectors in (\ref{eq:defkl}).  
\end{rem} 
\begin{rem}
Taking into account the preceding remark, one can see Lemma \ref{lem:te} as the generalisation of our Baxterisation formula to the mixed situation, namely for $\check R^{(k,l)}(u)$ for any $k,l$ (the condition $k\leq l$ is only for convenience, the situation is symmetric in $k$ and $l$). This generalised Baxterisation formula lives in the Hecke algebra $H_{k+l}(q)$. More precisely, it lives in the subspace $E^{(k,l)}:=P^{(k,l)}H_{k+l}(q)P^{(l,k)}$ which is not a subalgebra if $k\neq l$. However, one has a natural bilinear ``product'':
\[E^{(k,l)}\times E^{(l,k)}\to P^{(k,l)}H_{k+l}(q)P^{(k,l)}\ .\]
The subspace $P^{(k,l)}H_{k+l}(q)P^{(k,l)}$ is now a subalgebra which is also a fused Hecke algebra as defined in \cite{CPdA}. If $l=1$ (and $k$ arbitrary) it admits as a quotient a similar algebra where the Hecke algebra is replaced by the Temperley--Lie algebra. This is called in the literature the seam algebra (see for example \cite{LSA}) and thus one can see a particular case of (\ref{eq:lem}) as a Baxterisation formula for the seam algebra.
\end{rem}

\end{document}